\documentclass[final, 3p, times]{elsarticle}

\usepackage{amssymb}
\usepackage{amsmath}
\usepackage{amsfonts}
\usepackage{lmodern} 

\usepackage{graphicx}
\usepackage{hyperref}
\usepackage{verbatim}
\usepackage{algpseudocode}
\usepackage{algorithm}
\usepackage{multicol}
\usepackage{booktabs}
\usepackage{siunitx}

\usepackage{subcaption}

\newcounter{bla}

\journal{Computer Physics Communications}

\def\tr{\operatorname{tr}} 
\def\sym{\operatorname{sym}} 
\def\grad{\operatorname{grad}} 

\def\bg{{\mathbf{g}}}
\def\bk{{\mathbf{k}}}
\def\br{{\mathbf{r}}}

\def\bR{{\mathbf{R}}}

\def\bA{{\mathbf{A}}}

\def\bC{{\mathbf{C}}}

\def\bF{{\mathbf{F}}}

\def\bH{{\mathbf{H}}}

\def\bX{{\mathbf{X}}}

\def\dd{{\mathop{}\!\textrm{d}}}
\def\dr{{\dd\br\,}}

\def\ext{{\textrm{ext}}}
\def\H{{\textrm{H}}}

\def\xc{{\textrm{xc}}}

\def\KS{\textrm{KS}}

\def\half{\frac{1}{2}}

\begin{document}

\begin{frontmatter}



\title{
Direct minimization on the complex Stiefel 
manifold in Kohn-Sham density functional theory for finite and extended systems}


\author[a]{Kai Luo\corref{author}}
\author[a]{Tingguang Wang}
\author[b]{Xinguo Ren}

\cortext[author] {Corresponding author.\\\textit{E-mail address:} kluo@njust.edu.cn}
\address[a]{Department of Applied Physics, School of Physics, Nanjing University of Science and Technology, Nanjing 210094, China}
\address[b]{Institute of Physics, Chinese Academy of Sciences, Beijing 100190, China}
\begin{abstract}

Direct minimization method on the complex Stiefel manifold in Kohn-Sham density functional theory  
is formulated to treat both finite and extended systems in a unified manner. 
This formulation is well-suited for scenarios where straightforward iterative diagonalization becomes challenging, 
especially when the Aufbau principle is not applicable.
We present the theoretical foundation and numerical implementation of the  Riemannian conjugate gradient (RCG) 
  within a localized non-orthogonal basis set. 
Riemannian Broyden-Fletcher-Goldfarb-Shanno (RBFGS) method
is tentatively implemented.
Extensive testing compares the performance of the proposed methods 
 and highlights  that the quasi-Newton method is more efficient. 
However, for extended systems, the computational time  required
grows rapidly with respect to the number of $\bk$-points.

\end{abstract}


\end{frontmatter}




\begin{small}
\end{small}

\section{Introduction}
\label{sec:introduction}
Density functional theory (DFT) stands as a highly utilized method for simulating a wide range of physical systems,
including atoms, molecules, clusters, solids, and other complex forms of matter. 
This popularity stems from its exceptional balance between accuracy and computational efficiency. 
The ingenious implementation of Kohn-Sham (KS) theory within DFT precisely addresses the non-interacting 
kinetic energy through the solution of a one-body Schrödinger equation, incorporating an effective potential 
inclusive of exchange-correlation (xc) effects \cite{Hohenberg1964,Kohn1965}. 

Presently, the self-consistent field (SCF) algorithm derived from the first-order 
necessary optimality condition
is the prevailing method for tackling the Kohn-Sham problem. 
It revolves around identifying eigenvalues and corresponding eigenvectors within 
the occupied space of the Hamiltonian matrix. 
However, while widely adopted, this approach is susceptible to convergence issues. 
A well-designed density update scheme is thus indispensable for achieving convergence within a reasonable number of iterations \cite{Broyden1965,Pulay1980}. 
 Meanwhile, the converged solution may occasionally be a saddle point of the energy surface rather than a minimum \cite{Vaucher2017}. 

As an alternative, the Kohn-Sham problem can be treated as an optimization problem. 
One approach is the direct minimization of the Kohn-Sham energy functional with respect to the electronic degrees of freedom. 
This method requires ensuring that the orbitals remain orthonormal, thereby framing the task as a constrained optimization problem. 
This can be achieved by applying 
explicit orthonormalization, such as 
Gram-Schmidt or QR orthonormalization to the updated orbitals after each iteration \cite{Gillan1989,Payne1992}.
This constrained problem can also be reformulated into an unconstrained optimization problem using 
penalty function methods\cite{Nocedal_Wright,Bertsekas2014,Wen2016,Xiao2021}  or 
augmented Lagrangian (ALM) methods \cite{Nocedal_Wright,Hestenes1969,Powell1969, Bertsekas2014,Gao2019,Gao2022}. 
However, the smoothness requirement for penalty functions necessitates that the original objective function has high-order smoothness. 
In practice, calculating the exact gradients of these penalty functions for non-convex problem is often computationally expensive, 
and selecting appropriate penalty parameters can be challenging.
Recent parallelizable frameworks within the ALM method demonstrated effectiveness and high scalability, 
showing promise in electronic structure calculations\cite{Gao2019,Gao2022}.

The constraint can also be fulfilled by 
a unitary transformation matrix, which is applied to 
a set of orthonormal reference orbitals and then optimized.
Using the exponential transformation with a skew-Hermitian matrix exponent in a compact basis set
such as linear combination of atomic orbitals (LCAO), it has been shown that good performance can be achieved 
compared to the SCF method for both finite and extended systems
\cite{Douady1980,Rico1983a,Rico1983b, Head-Gordon1988, Ismail-Beigi2000,Voorhis2002, VandeVondele2003,Freysoldt2009,Ivanov2021}.
However, for non-compact basis functions of size $M$ (e.g. plane waves),  
computing the exponential of these matrices typically 
scales as  $\mathcal{O}(M^3)$ and thus making it computationally expensive.

In recent years, considerable progress has been made in the area of Riemannian 
optimization for the electronic structure theory. 
The nonconvex problem can be converted to a geodesically convex problem in 
a curved space. 
Since the Kohn-Sham energy is defined on a curved space (Riemannian manifold),
the optimization has to take the curvature into account\cite{Edelman1998,Absil2008_book}.
The extension of unconstrained optimization from Euclidean spaces leads to Riemannian optimization. 
 In addition to its effective utilization in various classical optimization problems with geometric restrictions, 
 Riemannian optimization has proven highly beneficial in electronic structure computations 
 \cite{Raczkowski2001,Bylaska2006, Wen2013, Zhang2014, Jiang2015, Baarman2016, Dai2017}.
 In these works by Wen et al. \cite{Wen2013}, Zhang et al.\cite{Zhang2014}, and 
 Dai et al. \cite{Dai2017}, finite systems were treated.  Ref. \cite{Dai2017} did not address periodic systems.  As a result, the underlying manifold is thus real in principle and only 
 when the basis (e.g. planewaves) is itself complex does the manifold become complex.
 However, in the periodic systems, complexity is required due to the intrinsic complexity of 
 the Bloch states. 
 We note that a recent 2023 paper\cite{Dai2023}, which is motivated by metallic systems and employs a plane-wave basis, uses Bloch-periodic boundary conditions for periodic calculations, as we do here; whereas our formulation is in terms of a general, non-orthogonal basis rather than a global, orthonormal one as in Ref. \cite{Dai2023}.

Despite of enormous success of the Kohn-Sham density functional theory, 
it finds difficulty in handling systems with strong correlations, such as 
Mott insulators. One promising theory for this matter going beyond KSDFT is the 
reduced density matrix functional theory (RDMFT) \cite{Sharma2008,Wang2022},
where the traditional iterative diagonalization meets difficulty 
and the Aufbau principle is not applicable. 
The orthogonality constraint for the natural orbitals (eigenfunctions of the 
one-particle reduced density matrix (1RDM)), 
 can be easily integrated in the Stiefel manifold.

In this study, we introduce a unified formulation adaptable to any basis for both 
finite and extended systems within the manifold minimization method.
An implementation based on inexact line search of the conjugate gradient (CG) 
(and tentatively Broyden-Fletcher-Goldfarb-
Shanno (BFGS)),
for the Kohn-Sham problem is provided within a non-orthogonal local basis set. 
By incorporating fractional occupation, both metallic and degenerate (or nearly degenerate) 
systems can be treated, which is only possible in the Stiefel manifold 
but not in the Grassmann manifold. 
Their performances on 
finite systems and extended systems are compared against the standard SCF method 
and discussed. 
Its success on the Kohn-Sham problem lays a solid foundation for 
the ongoing development of RDMFT and other theories that require non-idempotent 
density matrix.



\section{Theory Formulation}
\label{sec:theory}

\subsection{Notations}
\label{sub:notation}
For a complex matrix $A \in \mathbb{C}^{m\times n}$, matrices $A^\dagger$ and $A^{-1}$ denote 
the complex conjugate transpose and  inverse of $A$, respectively. 
For a vector $d \in \mathbb{C}^n$, the operator $\operatorname{Diag}(d)$ returns 
a square diagonal matrix in $\mathbb{C}^{n \times n}$ with the elements of $d$ 
on the main diagonal, while conversely $\operatorname{diag}(A)$ returns the vector 
in $\mathbb{C}^n$ containing the main diagonal elements of the square matrix 
$A \in \mathbb{C}^{n \times n}$.
$I_p$ is an identity matrix of size $p\times p$.
The symmetrized matrix of a square matrix $A$ is denoted as $\sym(A) = (A + A^\dagger) / 2$.
The trace of $A$, i.e., the sum of the elements on the main diagonal of a square matrix 
$A \in \mathbb{C}^{n \times n}$, is denoted by $\tr(A)$. 
The Frobenius inner product in Euclidean space for matrices $A, B \in \mathbb{C}^{m \times n}$ is 
defined as $\langle A, B\rangle_e=\tr\left(A^\dagger B\right)$, and 
the corresponding Frobenius norm $\|\cdot\|_F$ is given by 
$\|X\|_F=\langle A, A\rangle^{1 / 2}=\left(\sum_{i, j}\left|A_{i j}\right|^2\right)^{1 / 2}$.
For a $k$ indexed matrix $A_k$,  the boldface notation 
$\bA$ is to denote $(A_1, A_2, \ldots, A_K)$ and is of size $K$.
When we are specifically dealing with electronic structure problems, 
we may use indices such as $M, N$. Otherwise, $m,n,p$ will be used for a 
general matrix.

\subsection{Continuous Kohn-Sham DFT Model}
In Kohn-Sham density functional theory, the central quantity in the variational principle is the energy functional
\begin{equation}
\begin{aligned}
E_{\KS} [ \{ \psi_i(\br) \}] = & -\frac{1}{2} \sum_i f_i \int \dr \psi_i^*(\br) \nabla^2 \psi_i(\br) \\
+ E_{\H}[n] +  & E_{\xc} [n] +   \int \dr n(\br) v_{\ext}(\br) ,
\end{aligned} 
\label{eq:energy_functional}
\end{equation}
where the Hartree energy is given by
\begin{equation}
    E_{\H}[n] = \half \int\!\!\int \dr \dr' \frac{n(\br) n(\br')}{ | \br -\br'|} ,
\end{equation}
and $E_{\xc}[n]$ is the exchange-correlation energy functional, which has to be approximated in practice.

Here and throughout this article we work in atomic units and 
therefore have set $\hbar = m_e = e = 1$, where $m_e$ is the
electron mass and $e$ is the charge of the proton. 
 $v_{\ext}$ is the external potential for electron-nuclei interaction.

The electron density $n(\br)$ is the sum of the squared norm of 
the Kohn–Sham wave functions $\psi_i(\br)$ weighted by the smearing function 
$f(\epsilon, \mu)$, (e.g. the Fermi–Dirac distribution or the Gaussian smearing function)
\begin{equation}
    n(\br) = \sum_{i}^{\infty} f(\epsilon_i, \mu) \left|\psi_i(\br)\right|^2 .
    \label{eq:density}
\end{equation}
The chemical potential $\mu$ is chosen such that the total number of electrons is $N_e$.
In many density-matrix based formulations, it is useful to have the 
single-particle density matrix 
\begin{equation}
    \gamma(\br, \br') = \sum_{i=1}^{\infty} f(\epsilon_i, \mu) \psi_i^*(\br) \psi_i(\br') ,
\end{equation}
whose diagonal is the electron density
\begin{equation}
    n(\br) = \gamma(\br, \br) .
\end{equation}

Minimization of the energy functional subject to the orthogonality condition 
\begin{equation}
    \int \psi_i^*(\br) \psi_j (\br) = \delta_{ij}, 
\end{equation}
leads to the Kohn-Sham equation, 
\begin{eqnarray}
h \psi_i (\br)  & = & \epsilon_i \psi_i(\br) \label{eq:KSequation}\\
\nabla^2 v_\H(\br) & = &4 \pi n(\br)  \label{eq:Poisson}\\
v_{\xc}(\br) & = & \frac{\delta E_{\xc}[n]}{\delta n(\br)}
\end{eqnarray}
where the single-particle Hamiltonian is $h =-\frac{1}{2} \nabla^2+v_{\ext}(\br)+v_\H(\br)+v_{\xc}(\br) $.
The Hartree potential $v_{\H}$ may be obtained by solving the Poisson equation (see Eq.~\ref{eq:Poisson}).  
The dimension of the Hamiltonian matrix $H_{ij} \equiv \langle \psi_i | h |\psi_j \rangle $ is 
the size of the basis functions $M$.  However, due to the fast decaying property of the
 smearing function, only $N$ lowest eigenstates are needed.
Typically $N $ is  smaller than $M$ by a few orders of magnitude, especially for the case of 
non-compact basis, such as the plane-wave basis. 
Diagonalization of the Hamiltonian gives eigenvalues $\epsilon_i$ which are arranged from 
the smallest to the largest as 
$\epsilon_1 \leq \epsilon_2 \leq \dots \leq \epsilon_{N}$. 
This equation has to be solved iteratively due to the orbital dependence of the
 single-particle effective potential.  

\subsection{Matrix formulation}
The continuous Kohn-Sham model can be conveniently expressed in matrix notations 
and solved on a computer. To unify the treatment of both
finite  and extended systems, we explicitly include the $\bk$ dependence in the formulation. 
For each wave-vector $\bk$ in the 1st Brillouin zone, due to symmetry, there is a weight $\omega_{\bk}$ associated with it 
according to the space-group of the underlying structure.
Normally,  Kohn-Sham eigenstates are Bloch orbitals which can be
 expanded in terms of a possibly non-orthogonal basis functions 
 $\{\chi_{\mu\bk} \}$ 
of size $M$, 
\begin{equation}
    \psi_{i\bk } (\br) = \sum_{\mu=1}^{M} C_{\mu i\bk} \chi_{\mu \bk}(\br) \,.
\end{equation}
Here $i$ and $\bk$ are the band index and Bloch wave vector. The expansion coefficients $C_{\mu i\bk} $ can be regarded as a matrix whose $i$th column 
contains the expansion coefficients of the $i$th wave function indexed by $\bk$. 
For a $\Gamma$-point calculation in real basis, real coefficients can be used for memory saving and speedup. 
To keep it
general, we have 
$\bC \in \mathbb{C}^{M\times N \times K}$ (bold symbol for this size, see notations in Section \ref{sub:notation} ). Popular choices of basis functions are plane waves \cite{Payne1992}, Gaussian orbitals \cite{Davidson1986,Wilson1987}, 
(numerical) LCAO \cite{Li2016}, multiresolution analyses\cite{Arias1999}, or finite-difference/finite-element real-space grids 
\cite{Chelikowsky1994,Motamarri2020}. 

Within a local basis set, we might express the basis functions 
as 
\begin{equation}
    \chi_{\mu \bk}(\br)=\frac{1}{\sqrt{N}} \sum_{\bR} \exp (i \bk \cdot \bR) 
    \phi_\mu (\br-\tau_\mu -\bR )
\end{equation}
where $ \phi_\mu(\br-\tau_\mu -\bR)$ are the atomic orbitals centering 
on an atom in the unit cell $\bR$.
The index $\mu$ enumerates the atomic orbitals.  


The total density matrix is the sum 
\begin{equation}
    P = \sum_{\bk} P_{\bk},
\end{equation}
where the density matrix $P_{\bk}$ in the matrix representation is
\begin{equation}
    P_{\bk} = \omega_{\bk} C_{\bk} F_{\bk} C_{\bk}^{\dagger}
        \label{eq:density_matrix}
\end{equation}
with occupation matrix elements $F_{ij\bk} = f(\epsilon_{i\bk}, \mu) \delta_{ij}$.
The charge density can be expressed as the diagonal of the density matrix,
\begin{equation}
    n(\br) = \operatorname{diag} ( \langle \br| P | \br'\rangle ),
\end{equation}
in which $P$ is evaluated on grid points.
The chemical potential $\mu$ can be determined by satisfying $N_e = \int\!\dr n(\br)$, with 
$N_e$ electrons in the unit cell.

The Kohn-Sham equation (see Eq.~\ref{eq:KSequation}) 
can be cast into a generalized matrix eigenvalue problem,
\begin{equation}
    H_{\bk}  C_{\bk} = E_{\bk} S_{\bk}  C_{\bk} ,
    \label{eq:generalized_eigen}
\end{equation}
where $H_{\bk}$, $S_{\bk}$, and $C_{\bk}$ are the Hamiltonian matrix, overlap matrix and 
eigenvectors at a given $\bk$-point, respectively. 
The energy matrix $E_{\bk}$ is a diagonal matrix with elements $E_{ij\bk}= \epsilon_{i\bk} \delta_{ij}$. 
In the case of norm-conserving pseudopotentials, the external potential can be split into local part $v_\alpha^{L}$ and nonlocal part $v_\alpha^{NL}$, $v_{\ext, \alpha} =v_\alpha^{L} +  v_\alpha^{NL}$.  
The nonlocality of pseudopotential $v^{NL}$ is included via the standard nonlocal projectors \cite{Kleinman1982},
\begin{equation}
    v_\alpha^{N L}=\sum_{l=0}^{l_{\max }} \sum_{m=-l}^{l} \sum_{n=1}^{n_{\max }}\left|\chi_{\alpha l m n}\right\rangle\left\langle\chi_{\alpha l m n}\right|
\end{equation}
where $\left|\chi_{\alpha lmn }\right\rangle$ are non-local projectors.
 $l$ and $m$ are the azimuthal and magnetic quantum numbers, respectively, and n is the multiplicity of projectors.
$l_{\max }$ and $n_{\max }$ are the maximal angular momentum and the maximal multiplicity of projectors for each angular momentum channel.

The Hamiltonian matrix $H_{\bk}$ and  the overlap matrix $S_{\bk}$ are both of size $M\times M$,
\begin{equation}
    H_{\alpha \beta \bk} = \int\!\!\dr \chi_{\alpha\bk}^*(\br) H_{\bk} \, \chi_{\beta\bk}(\br) \,
\end{equation}
and 
\begin{equation}
    S_{\alpha \beta \bk} = \int \!\!\dr \chi_{\alpha\bk}^*(\br) \, \chi_{\beta\bk}(\br).
\end{equation}
For orthonormal basis functions, such as the plane wave basis, $S_{\bk} = I_M$. In general, 
the overlap matrix $S_{\bk}$ is a symmetric positive definite matrix, which assures a Cholesky decomposition 
\begin{equation}
    S_{\bk} = U_{\bk}^\dagger U_{\bk}.
    \label{eq:cholesky_decomp}
\end{equation}
The orthogonal constraint imposed on the coefficient matrix reads
\begin{equation}
    C_{\bk}^\dagger S _{\bk} C_{\bk\dagger} = I_N .
    \label{eq:ortho_constraint}
\end{equation}
The total energy is thus a function of the coefficient matrix $\bC$,  
$E_{\KS}(\bC) $.

\subsection{Matrix optimization with orthogonality constraints}
As introduced above, instead of diagonalizing the Hamiltonian matrix, an alternative method is called the direct minimization, which was discussed in details in Ref. \cite{Payne1992}. 
The standard optimization problem with orthogonality constraints is 
\begin{equation}
\min_{X \in \mathbb{C}^{n \times p}} \mathcal{F} (X), \text { such that } X^{\dagger} X= I_p,  
\end{equation}
where $\mathcal{F}(X): \mathbb{C}^{n \times p} \to \mathbb{R}$ is a differentiable function. 
Meanwhile, the KS minimization problem becomes 
\begin{equation}
 \min_{C_\bk \in \mathbb{C}^{M \times N}} E_{\KS}(\bC), 
  \text { such that } C_{\bk}^\dagger S_{\bk} C_{\bk} = I_N.
 \label{eq:KS_problem}
\end{equation}
It can be easily verified that  the KS problem can be adapted to the standard form,
with the auxiliary transformation matrix $U_{\bk}$ (see Eq.~\ref{eq:cholesky_decomp}), 
\begin{equation}
    X_{\bk} = U_{\bk} C_{\bk}  \quad \text{ or } \quad C_{\bk} = U_{\bk}^{-1} X_{\bk}.
    \label{eq:transform}
\end{equation}



The total energy can be written as a sum
\begin{equation}
    E_{\KS} =  E_b[P] + \Phi [n],
\end{equation}
where 
\begin{equation}
    E_b [P]= \tr(\sum_{\bk} P_{\bk} H_{\bk}),
    \label{eq:band_energy}
\end{equation}
is the  band energy and $\Phi[n]$ is a density-dependent quantity.
The derivative of the energy functional is essential in the optimization.
The energy variation can be computed with the variation of the density matrix 
\begin{eqnarray}
        \dd E_{\KS} &=& \tr(\sum_{\bk} H_{\bk} \dd P_{\bk})
\end{eqnarray}
where, substituting Eq.~\ref{eq:transform} into Eq.~\ref{eq:density_matrix}, 
\begin{eqnarray}
   \dd P_{\bk} &=& \omega_\bk \left[
  \dd C_{\bk} \, F_{\bk}  \,C_{\bk}^\dagger  
   +  C_{\bk} \, F_{\bk} \, \dd C_{\bk}^\dagger   \right] \nonumber \\
 &=&  \omega_\bk \left[
    U_{\bk}^{-1} \left( \dd X_{\bk} \, F_{\bk}    \, X_{\bk}^\dagger + 
  X_{\bk} \, F_{\bk}    \, \dd X_{\bk}^\dagger \right)  (U_{\bk}^{-1})^{\dagger} \right]  \,.
\end{eqnarray}
Therefore, the derivative of the energy functional can be derived as follows 
\cite{Ismail-Beigi2000}
\begin{equation}
      (\nabla E_{\KS})_{\bk} \equiv  \frac{\partial E_{\KS}}{\partial X_{\bk}^\dagger } = 
       \omega_\bk (U_{\bk}^{-1})^\dagger H_{\bk} U_{\bk}^{-1} X_{\bk} F_{\bk}, 
       \label{eq:gradient}
\end{equation}
following the definition
\begin{equation}
        \dd E_{\KS} = \tr \left[ \dd X_{\bk}^\dagger \left( \frac{\partial E_{\KS}}{\partial X_{\bk}^\dagger } \right) +
          \left( \frac{\partial E_{\KS}}{\partial X_{\bk}^\dagger } \right) ^\dagger \dd X_{\bk}
        \right]
\end{equation}
and the application of cyclicity of the trace $\tr(X Y) = \tr(Y X)$.


\subsection{Complex Stiefel manifold}    
Classical methods for unconstrained optimization in Euclidean space, 
such as steepest descent, conjugate gradient, quasi-Newton methods, and 
trust-region methods, can be generalized to optimization on Riemannian manifolds.
For the KS problem (\ref{eq:KS_problem}), the underlying manifold is a complex Stiefel manifold, 
which is the space of matrices defined as
\begin{equation}
    \mathrm{St}^{p}_{n} :=  \{ X \in \mathbb{C}^{n\times p}: X^\dagger X = I_p \}.
    \label{eq:stiefel_manifold}
\end{equation}
We denote the manifold as $\mathrm{St}$ for brevity. The Stiefel manifold may be 
embedded in the $np$-dimensional Euclidean space of $n$-by-$p$ matrices. 
When 
$p=1$, the Stiefel manifold reduces to a sphere, and when 
$p=n$, it corresponds to the group of orthogonal matrices, known as $O_n$.
At $X \in \mathrm{St}$, we have the tangent space 
\begin{equation}
    T_X \mathrm{St} = \left\{ Y = X B + Z \mid B^{\dagger} = -B,  Z^{\dagger} X = 0 \right \},
    \label{eq:tangent_manifold}
\end{equation}
where $  Y, Z \in \mathbb{C}^{n\times p}, B \in \mathbb{C}^{p\times p} $. 
Here, 
$B$ is a skew-Hermitian matrix  and $Z$ is a matrix orthogonal to $X$.

The orthogonal projection of any vector $V\in  \mathbb{C}^{n\times p}$ onto 
 $T_X \mathrm{St}$ is 
\begin{equation}
    \pi_{X} (V) = V- X \sym (X ^{\dagger} V).
    \label{eq:projection}
\end{equation}
There are two commonly used metrics for the tangent space:
the Euclidean metric $\langle U , V\rangle_{X}^{e} = \tr(U^ \dagger V)$ and the canonical metric 
\begin{equation}
    \langle U, V\rangle_{X}^{c} = \tr \left[ U^{\dagger} \left( I_n - \half X X ^{\dagger} \right) V\right]
    \label{eq:canonical_metric}
\end{equation}
where $U , V\in T_X \mathrm{St} $.

In Riemannian optimization, two fundamental components are needed.
The first component is the ``retraction", which smoothly maps a point from the 
tangent space to the manifold. 
In notations, a mapping $\mathcal{R}$ from the tangent space $T \mathrm{St}$ into $\mathrm{St}$ is a retraction, which satisfies 
  \begin{subequations}
    \begin{align}
       & \mathcal{R}_X (0) = X, \forall X \in \mathrm{St} ,
       \\
       & \frac{d}{d t} \mathcal{R}_X(t Z )\mid _{t=0} = Z, \forall  X \in \mathrm{St}, \forall Z \in T_X \mathrm{St}.
    \end{align}
\end{subequations}  
Common retractions for Stiefel manifold include the exponential mapping 
\begin{equation}
    \mathcal{R}^{exp}_X(U)=\left(
X \quad U
\right)\left(\exp \left(\begin{array}{cc}
A & -S \\
I_p & A
\end{array}\right)\right)\binom{I_p}{0} \exp (-A)
\end{equation}
where $X \in \mathrm{St}, U \in T_{X} \mathrm{St}, A = X^{\dagger} U$, 
and $S = U^{\dagger} U$ \cite{Edelman1998}.
This retraction requires geodesics along the manifold, 
where matrix exponential is required and thus computationally expensive. 
In contrast, projection-like retractions such as the QR decomposition 
can be viewed as the first-order approximations 
to the exponential one, which is preferred in many practical applications.
The QR decomposition retraction is 
\begin{equation}
        \mathcal{R}_X(U)=\operatorname{qf}(X+U) \label{eq:qr_decomp}
\end{equation}
where $\operatorname{qf}(\cdot)$ denotes the $Q$ factor of the QR decomposition with non-negative elements on the
diagonal of $R$. The polar decomposition 
\begin{equation}
    \mathcal{R}_X(U)=(X+U)\left(I_p+U^\dagger U\right)^{-1 / 2} \label{eq:polar_decomp}
\end{equation}
 is another common retraction choice, which is second-order. 

The second component is the ``vector transport",
which transfers a vector from the tangent space of an adjacent point 
to the same tangent space at a given point. 
It is a computationally affordable approximation to the ``parallel transport".
This is essential in the 
optimization approaches such as the conjugate gradient method,
because otherwise vectors from different tangent spaces are not directly computable. 
The vector transport by projection is denoted by $\mathcal{T}^P$, as in Eq.~\ref{eq:projection},
\begin{equation}
    \mathcal{T}^P_U  (V)= V- Y \sym ( Y^\dagger V) 
    \label{eq:vector_transport}
\end{equation}
where $U, V\in T_{X} \mathrm{St}, Y = \mathcal{R}_{X}(U) $, and $\mathcal{R}$ 
is the associated retraction. 
Alternatively, the vector transport by 
differentiated retraction $\mathcal{T}_U^R(V)$
could be used accordingly \cite{Zhu2017}.

\subsection{Riemannian conjugate gradient methods}
 
Conjugate gradient methods offer significant advantages by efficiently handling the curvature and 
geometric structure of the manifold, requiring low memory, and achieving faster convergence. 
They avoid the need for matrix inversions and are adaptable with retraction methods, 
making them versatile and powerful tools for manifold-based optimization problems in various scientific and engineering applications.

Similar to the Euclidean case, the Riemannian conjugate gradient (RCG) methods require the essential ingredient, 
the Riemannian gradient $g = \grad f$. In the Stiefel manifold, it can be computed with the Euclidean gradient 
$\nabla f$
\begin{equation}
    \grad f = \nabla f - X \left( \nabla f \right)^\dagger X.
    \label{eq:Riemannian_gradient}
\end{equation}
Keeping the conjugacy, the new search direction $d_{k+1}$ is computed as 
\begin{equation}
    d_{k+1} = - g_{k+1} + \beta_{k+1}    \mathcal{T}_{\alpha_k d_{k}} (d_k)
\end{equation}
where $g_{k}$ denotes the gradient at iteration $k$ and 
$\mathcal{T}_{\alpha_k d_{k}} (d_k)$ in this work is the projection base formula in 
Eq.~\ref{eq:vector_transport}.
In RCG methods, the parameter $\beta_{k+1}$ 
for the conjugate gradient direction in each iteration can take on various forms.
 To facilitate this, it is often convenient to define the quantity:
\begin{equation}
    y_{k+1} = g_{k+1} - \mathcal{T}_{\alpha_k d_{k}} ( g_{k} )
\end{equation}
With this,  schemes by Fletcher-Reeves \cite{FletcherReeves1964}, Polak-Ribi\`ere \cite{PolakRibiere1969} and Polyak\cite{Polyak1969}, 
Dai-Yuan\cite{Dai1999}, Hestenes-Stiefel\cite{Hestenes1952}, Liu-Storey\cite{Liu1991}, Hager-Zhang \cite{HagerZhang2005} 
are commonly adopted algorithms. 
Some examples of these adapted forms are listed below.
\begin{subequations}
\begin{align}
\beta_{k+1}^{\mathrm{FR}} &=\frac{\langle g_{k+1}, g_{k+1} \rangle_{X_{k+1}}}{\langle g_{k}, g_{k} \rangle_{X_{k}}}, \\
\beta_{k+1}^{\mathrm{PR-P}} &=\frac{ \langle g_{k+1}, y_{k+1} \rangle_{X_{k+1}}}{\langle g_{k}, g_{k} \rangle_{X_{k}}}, \\
\beta_{k+1}^{\mathrm{DY}} &=\frac{\langle g_{k+1}, g_{k+1} \rangle_{X_{k+1}}}{\langle y_{k+1}, \mathcal{T}_{\alpha_k d_k }(d_k) \rangle_{X_{k+1}}}, \\
\beta_{k+1}^{\mathrm{HS}} &=\frac{\langle g_{k+1}, y_{k+1} \rangle_{X_{k+1}}}{\langle y_{k+1}, \mathcal{T}_{\alpha_k d_k }(d_k) \rangle_{X_{k+1}}}.
\end{align}
\label{eqs:cg_algorithms}
\end{subequations}
The canonical metric in Eq.~\ref{eq:canonical_metric} is adopted in computing the vector product, 
e.g. $\langle y_{k+1}, \mathcal{T}_{\alpha_k d_k }(d_k) \rangle_{X_{k+1}}$ and hence all superscripts $e$ is omitted. 
With these, the RCG method is summarized in Algorithm \ref{algs:cg}. In this algorithm, we give an example of QR decomposition retraction in the language of linear algebra operations.  For $\mathcal{R}_{X_k}(\alpha_k d_k)= \operatorname{qf}(X_k + \alpha_k d_k)$,
$\operatorname{qf}(X_k + \alpha_k d_k)$ is the Q factor of the QR decomposition of $X_k + \alpha_k d_k$.
For other retractions and vector transports, one needs to apply the corresponding linear algebra operations in Table  \ref{tab:linear_algebra}.

For multiple $\bk$-points, all quantities are 
indexed by  $ k_i = 1, 2, \cdots, K $, where $K$ is its total size. 
Therefore, the concept of the product of manifolds naturally fits into the description. 
A product of Stiefel manifolds is denoted by $\mathrm{St}=\mathrm{St}_1 \times 
\mathrm{St}_2 \times \cdots \times \mathrm{St}_K$, 
where $\mathrm{St}_i$ is a sub-manifold. 
An element $\bX$ in $\mathrm{St}$ is denoted by 
\begin{equation}
    \bX=\left(X_1^T, X_2^T, \cdots, X_K^T\right)^T, 
\end{equation}
where $X_i \in \mathrm{St}_i$.  The tangent space of $\mathrm{St}$ is
\begin{equation}
T_X \mathrm{St}=T_{X_1} \mathrm{St}_1 \times T_{X_2} 
\mathrm{St}_2 \times \cdots \times T_{X_K} \mathrm{St}_K.
\end{equation}
whose norm is thus the sum of all the norms of each sub-manifold.

To apply to the multiple $\bk$-points cases, one simply expands the dimension of 
 the pertinent $X$, by stacking $K$ copies of $X$,
  each of the same size.  In the algorithm, one needs to
  adapt $X$ and $g$ into bold symbols $\bX$ and $\bg$ (see above).  For each $\bk$-point
   taking care of the orthogonal constraint $X^\dagger_{k_i} X_{k_i} =  I_p$,
   the related operations in the algorithm has to be performed within the corresponding submanifold.
The  modification to the norm is to use the proper norm for the product of manifolds (e.g. 
the maximum norm in Ref. \cite{Dai2023})
when computing the CG parameters. 

\begin{table}[h!]
    \centering
    \caption{Example of key linear algebra operations for Riemannian optimization on the Stiefel manifold. Note, $Y = \mathcal{R}_X(U)$ in the vector transport formula. }
    \label{tab:linear_algebra}
    \begin{tabular}{@{}llc@{}}
    \toprule
    \textbf{Operation} & \textbf{Linear Algebra Formula} & \textbf{Key LAPACK Routine} \\ \midrule
    Euclidean metric & $\langle U,V \rangle^e_X = \tr[U^\dagger V]$ & ZGEMM \\
    Canonical metric & $\langle U,V \rangle^c_X = \tr[U^\dagger(I - \frac{1}{2}XX^\dagger)V]$ & ZGEMM \\
    Riemannian gradient & $\grad f(X) = \nabla f - X (\nabla f)^\dagger X$ & ZGEMM \\
    QR retraction & $\mathcal{R}_X(U) = \operatorname{qf}(X+U)$ & ZGEQRF \\
    Polar retraction & $\mathcal{R}_X(U) = (X+U)(I_p+U^\dagger U)^{-1/2}$ & ZHEEV \\
    Vector transport by projection & $\mathcal{T}_U (V) = V - Y \sym(Y^\dagger V)$ & ZGEMM \\
    Tangent space projection & $\pi_X(V) = V - X \sym(X^\dagger V)$ & ZGEMM \\
    \bottomrule
    \end{tabular}
\end{table}

\begin{algorithm}
    \caption{Conjugate gradient method for minimizing $f(X)$ on the 
    Stiefel manifold. }
    \label{algs:cg}
    \begin{algorithmic}[1]
    \State Initialization: choose $X_0 \in \mathrm{St}$,  $\epsilon_g, \epsilon_f > 0$,
    $k_{\max}$, 
      $g_0 = \grad f(X_0)$,
     initial search direction $d_0 = g_0$ 
    \While{$\|g_k\| > \epsilon_g$ (or $|f_{k+1} - f_{k}| >  \epsilon_f$) \textbf{and} $k < k_{\max}$}
        \State Line search to find step size $\alpha_k > 0$, 
        and update the point $X_{k+1} \gets  \mathcal{R}_{X_k}( \alpha_k d_k) $ using Eq.~\ref{eq:qr_decomp}  
        \State Compute new Riemannian gradient $g_{k+1} \gets \grad f(X_{k+1})$ 
        using Eq.~\ref{eq:Riemannian_gradient} 
        and the conjugate direction parameter $\beta_{k+1}$ (e.g. using the FR scheme)
        \begin{equation*}
            \beta_{k+1} \gets \frac{\langle g_{k+1}, g_{k+1} \rangle _{X_{k+1}} } {\langle g_k, g_k \rangle _{X_{k}}}
        \end{equation*} 
        \State Compute a search direction as $d_{k+1} \gets - g_{k+1} + \beta_{k+1}    
        \mathcal{T}_{\alpha_k d_{k}} (d_k)$ using Eq.~\ref{eq:vector_transport} 
        \State $k \gets k + 1$
    \EndWhile
    \end{algorithmic}
    \end{algorithm}
More graphically, we represent this algorithm in the following flowchart.
\begin{figure}[htbp]
    \centering
    \includegraphics[width=0.80\textwidth]{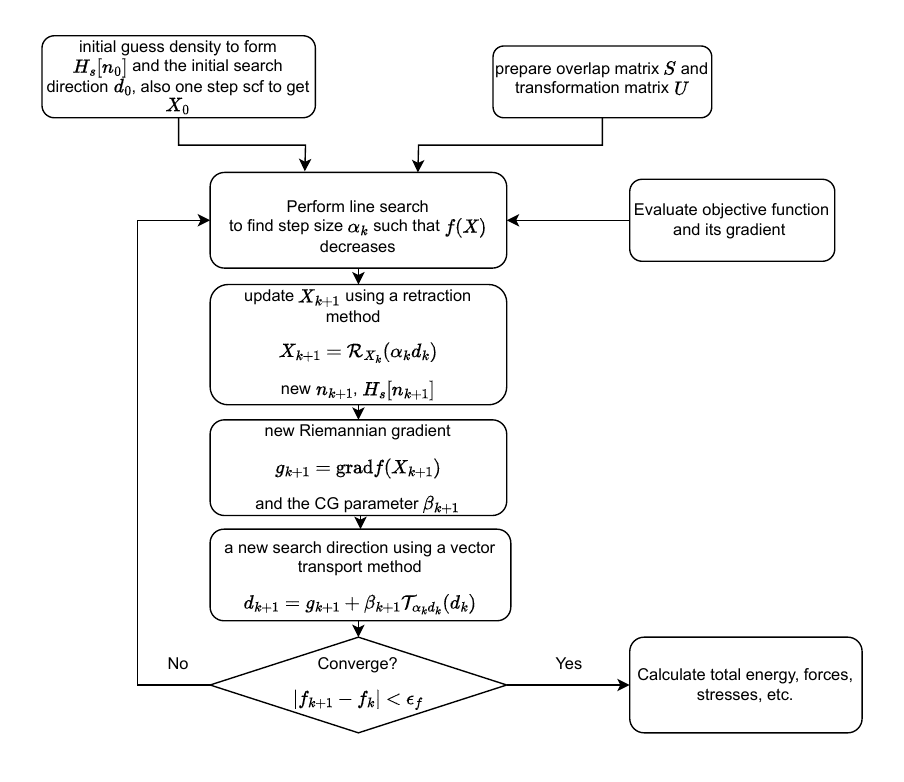}
    \caption{The flowchart for the RCG method. }
    \label{fig:flowchart}
\end{figure}
The most time consuming part is in the line search part 
where function and its gradient evaluations are required to find the 
right step size. In contrast, the standard SCF method is rather different (see Figure 
\ref{fig:SCF_flowchart} in \ref{sec:appendix_scf}). It requires direct (or iterative) diagonalization of the Hamiltonian matrix and a proper density mixing scheme.

\section{Implementation details}
\label{sec:implementation}
As a first step, we have implemented the conjugate gradient direct minimization on the complex Stiefel manifold algorithm 
within the open-source ABACUS software \cite{Chen2010,Li2016}, which uses norm-conserving pseudo potentials to 
describe the interactions between nuclear ions and 
valence electrons. 
Currently, we have only implemented the numerical atomic basis set calculations. 
The same algorithm can be easily adapted to the plane-wave basis, which 
ABACUS also supports.

Taking advantage of the modular structure, the whole algorithm is integrated as a new inherited solver. 
In this solver, typical conjugate gradient schemes \ref{eqs:cg_algorithms} can be chosen via 
variables within the input. 
Choices of retraction defaults to the projection type and the vector transport by projection $\mathcal{T}^P$ is used. 
The step length is chosen such that it satisfies the strong Wolfe conditions \cite{Zhu2017}
\begin{equation}
f\left(\mathcal{R}_{X_k}\left(\alpha_k d_k\right)\right) \leq f\left(X_k\right)+c_1 \alpha_k\left\langle\nabla f\left(X_k\right), d_k\right\rangle_{X_k}
\end{equation}
and
\begin{equation}
\left|\left\langle\grad f\left(\mathcal{R}_{X_k}\left(\alpha_k d_k\right)\right), \mathcal{T}_{\alpha_k d_k }(d_k) 
\right\rangle_{\mathcal{R}_{X_k}\left(\alpha_k d_k\right)}\right| \leq c_2\left|\left\langle\grad f\left(X_k\right), d_k\right\rangle_{X_k}\right|
\label{eq:strong_wolfe}
\end{equation}
where $0 < c_1 < c_2 < 1$. Default values $c_1 = 10^{-4}, c_2 = 0.9$ are used.
The initial step length $\alpha_0 = 1.0 $ is used as default and can be modified in the input. 
The line search begins with a trial estimate $\alpha^{t}$ 
, and keeps
increasing the step length until it finds either an acceptable step length or an interval that brackets the
desired step lengths. 
Once such an interval is established, the \textit{zoom} algorithm, combined with cubic interpolation, is employed to refine the step length. This bracketing process continues until an acceptable step length is determined \cite{Nocedal_Wright}.
We have not done any preconditioning to speed up the convergence yet.
An initial guess for the orbitals is taken to
be the eigenvectors of the Hamiltonian obtained from a superposition of atomic densities.
The implementation can be found in Ref.\cite{abacusdirectmin}.

\begin{figure}
    \centering
    \includegraphics[width=0.5\textwidth]{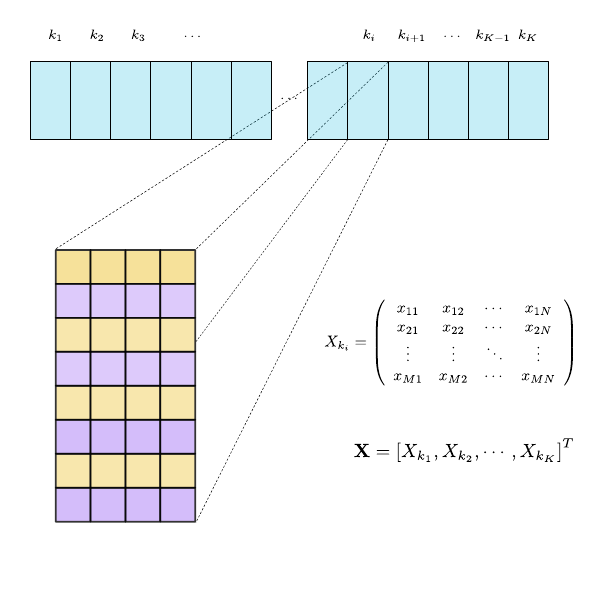}
    \caption{Data structure for $\bk$-dependent variable $X$ where each component 
    of the vector indexed by $k_i$ lives in a  sub-manifold. Each matrix $X_{k}$ is a 
    $M\times N$ complex matrix. Same structure applies to 
    $\bH$ and $\bF$.}
    \label{fig:matrix_structure}
\end{figure}

Due to the $\bk$ dependence in the orbitals $X_{\bk}$, the dimension of the one-electron wave-function 
is of size $M\times N \times K$, where $K$ denotes the number of irreducible $\bk$-points in the 
1st Brillouin zone. For this, 
we use the concept of product of Stiefel manifolds as a single entity.
Therefore, for each $\bk$, there 
is a sub-manifold associated with it. The data structure is sketched in Figure \ref{fig:matrix_structure}.
To assess the feasibility for varying $K$, the norm of $\bX$ is defined as the 
sum of norms in each sub-manifold divided by the number of sub-manifolds, $K$. 
In other words, we take the average. In this way, 
its norm $\|\bX\|$ should be close to 1 for varying $K$. 

The structure is very similar to a typical representation of wave-functions. What is needed in Algorithm \ref{algs:cg} is to empower the linear algebra operations to them.
This could easily be realized with some open-source linear algebra libraries, such as  \textit{armadillo} \cite{Armadillo} or \textit{Eigen}\cite{Eigen} in C++ language. In our implementation, we used a vector of size $K$ of  complex matrices in \textit{armadillo} to represent $X$. Each vector element is of size $M\times N$.
The linear algebra operations are performed independently within each sub-manifold.

To form $F_{\bk}$, the occupation number $f_{i\bk}$ can be computed with only the diagonal 
elements of $C^{\dagger}_{\bk} H_{\bk} C_{\bk}$.
The chemical potential $\mu$ is determined by bisection method 
according to a smearing scheme such as the Fermi-Dirac smearing or the Gaussian smearing. 

\section{Results}
\label{sec:results}

All calculations were performed on a workstation with an Intel(R)  Comet Lake Processor (at 3.80 GHz$\times$8, 16MB cache). The total number of cores is 8 and the total number of threads is 16. All codes were compiled with the Intel oneAPI compilers on Debian 12. To ensure fair comparisons, multi-threading was disabled, and only a single core was utilized for all computations.

\subsection{Test problems}
To test the RCG algorithm, we have applied the optimization procedure to two simple problems. 
The argument $X_{k} \in \mathbb{C}^{n \times p} $  in both problems are
subject to the orthogonality constraint
$X_{k}^{\dagger} X_{k} = I_p $. To mimic the $\bk$ dependence in the electronic 
structure theory for periodic systems, we have extended these with a $k$ dependence of minimum size 2, namely $K=2$. 

The first problem is the orthogonal Procrustes problem.
In this problem, the objective function is $f(X) = \| A X - B\|_F $  and its gradient can be computed analytically $\frac{\partial f}{\partial X^{\dagger}} = A X - B$. 
The second problem is the eigenvalue problem, whose objective function is 
$f(X) = -\half \tr \left( X^{\dagger} E X \right)$ and its gradient is 
$\frac{\partial f}{\partial X^{\dagger}} = -E X$. 
 For each $k$, $A_{k} \in \mathbb{C}^{m \times n},B_{k} \in \mathbb{C}^{m\times p}$, and  $E_{k} \in \mathbb{C}^{n \times n}$. 
 Setting random $A$ and $B = A I$, an initial guess is chosen as 
 the known solution $I$ plus a perturbed random deviation $X_0 = I + \epsilon P$, where $\epsilon$ is a small number and $P$ is a random matrix,
 the algorithm successfully finds the solution.
 Similarly, setting $E$ as a Hermitian matrix, the algorithm also delivers the right solution against the standard eigensolver.

\subsection{Molecules and Solids}
 To identify the advantages and disadvantages of the RCG method, 
 we performed single point ground state energy calculations for G2 data 
 set of small molecules\cite{Curtiss1997,Curtiss1998}, and a few simple bulk solids. 
 As a comparison, we have also provided a tentative implementation of the Riemannian BFGS (RBFGS) method 
 (see detailed algorithm in \ref{sec:appendix_bfgs}),
 which has to be considered preliminary (some more cases fail to converge). 
 Structure files were converted to ABACUS format using the 
 utility code \textit{atomkit}.  For 
 molecules, a cubic supercell of length 15.0 Angstrom with $\Gamma$-point 
 is always used and periodic boundary conditions are imposed.  
 For solids, Monkhorst-Pack k-point meshes were generated with a spacing of 0.06 in the unit of $2\pi/$\AA \; with \textit{atomkit} as well. 
 Norm-conserving pseudopotentials (optimized ONCV) \cite{Schlipf2015} were used to describe the electron-ion interactions. 
 The nonlocality of pseudopotential is included via the standard nonlocal projectors \cite{Kleinman1982}. 
 Double-$\zeta$ 
 plus polarization function (DZP) basis set and the Perdew-Burke-Ernzerhof (PBE) exchange-correlation functional were used\cite{PBE1996}. 
 The default SCF algorithm uses the Broyden density mixing 
 of dimension 8\cite{Johnson1988}, with mixing parameter 0.8 for spin-unpolarized calculations. 
 The convergence is reached when the relative density error, $\Delta \rho_R=\frac{1}{N_e} \int|\Delta \rho(r)| d^3 r$,
 is less than $10^{-6}$. By default, the Kerker preconditioner 
 is also turned on. The direct minimization stops when the change of the total energy 
 between consecutive iterations is less than  $5\times 10^{-9}$ a.u.
 Note here we use 
 $ | f_{k+1}  - f_{k} |<  \epsilon_f$ as the stopping criterion 
 in Algorithm \ref{algs:cg}. 
 After numerous tests, we found that the function tolerance $\epsilon_f$  is more effective in controlling the convergence than the gradient  tolerance $\epsilon_g$.

To begin with, we implemented four conjugate gradient schemes listed in Eq.~\ref{eqs:cg_algorithms}. 
For a test molecule acetonitrile $\mathrm{CH_{3}CN}$, the converged energy from the SCF calculation 
is taken as the minimum energy. The error in terms of iteration is shown on a log scale. 
It can be seen that all schemes have the super-linear convergence. 
For this case, Dai-Yuan and Polak-Rie\`ere-Polyak are slightly faster in reaching the minimum. 
It is also observed that Dai-Yuan runs faster than others in many cases. 
Therefore, all our calculations used Dai-Yuan scheme in computing $\beta_{k}$.
\begin{figure}[ht]
    \centering
    \includegraphics[width=0.6\textwidth]{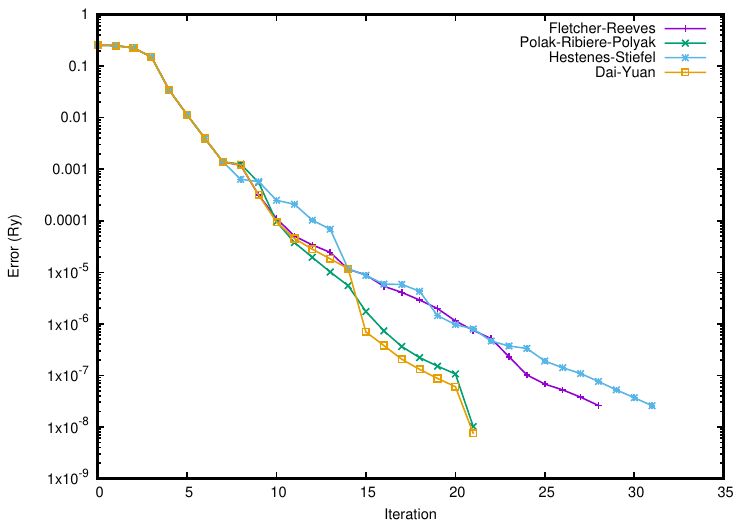}
    \caption{Convergence rate comparison among four common conjugate gradient schemes for RCG calculation on acetonitrile molecule, $\mathrm{CH_{3}CN}$. Error is taken with respect to the converged energy of SCF and is displayed on a log scale.  Among these schemes, the Dai-Yuan method consistently consumes the least time in most scenarios.}
    \label{fig:cg_converge}
\end{figure}
 
\begin{figure}[ht]
    \centering
    \includegraphics[width=0.6\textwidth]{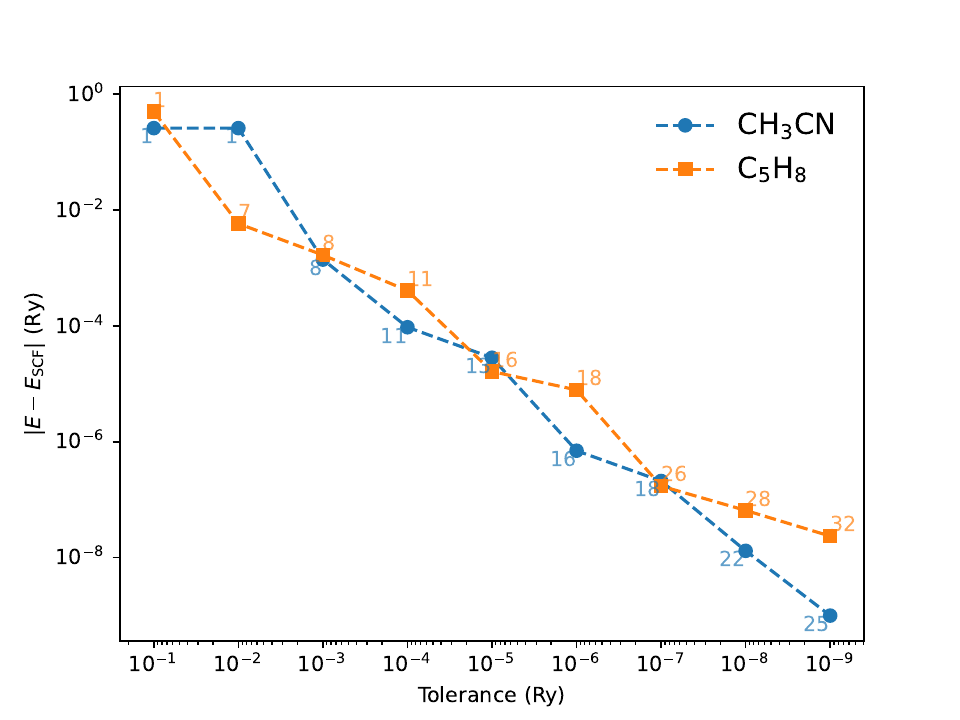}
    \caption{The error of the ground state energy with respect to the different tolerances for
    $\mathrm{CH_3CN}$ and $\mathrm{C_5H_8}$.
    $E_\mathrm{{SCF}}$ is the converged energy from SCF calculation.
    The number of iterations is shown beside the marker.}
    \label{fig:tol_error}
\end{figure}

Additionally, to examine how the error in the ground state energy changes 
based on different stopping criteria and tolerance levels in RCG, 
we present the results in Fig. \ref{fig:tol_error} 
for the acetonitrile ($\mathrm{CH_3CN}$) and spiropentane ($\mathrm{C_5H_8})$ molecules.
We observe that a smaller tolerance $\epsilon_f$ leads to a smaller error in the ground state energy, 
which aligns with our expectations. The number of iterations increases when the tolerance decreases, 
as anticipated. 

Calculations for all 148 molecules in the G2 dataset are successfully converged using the SCF method. 
However, for some molecules, the minimization process of the RCG method halted after 
the first few iterations, resulting in higher energies compared to the SCF calculations. 
This issue may stem from the initial gradient being too small due to the atomic density initialization.
 For nearly all the remaining 134 molecules, the RCG method yielded total energies with an error of 
 less than 0.003 \si{meV} (mostly 0.001 \si{meV}) compared to the SCF energies. 



The number of iterations of RCG is compared to that of  RBFGS. 
In Table \ref{tab:stats}, the statistics show that the RCG method has larger 
variations in the number of iterations. The average iterations 
of RBFGS is less than the RCG method. 
From these tests, the RBFGS method should be the method of choice.


\begin{table}[ht]
    \centering
    \begin{tabular}{|c|  c c | }
        \hline
        Iterations   & RCG & RBFGS \\
        \hline
        Avg.     & 26.5 & 16.1 \\
        Std.     & 13.6      &  4.2 \\
        Min    &    7   & 4     \\
        Max    & 84  & 29 \\
        \hline
    \end{tabular}
    \caption{Statistics for the total number of iterations for 
    RCG and RBFGS  methods for molecules. 
    Avg. and Std. stand for the mean and the standard deviation.
    }
  \label{tab:stats}
\end{table}

The Hamiltonian is updated in both real space and reciprocal space, a process that involves 
computationally expensive numerical grid integrals and Fourier transforms. 
This step is common to both the direct minimization method
and the iterative diagonalization method, resulting in similar computational costs per iteration.
During the line search, the RCG and RBFGS method may require a couple of calls in evaluating the 
objective function and its gradient. 
In contrast, for small system sizes, such as those in the G2 dataset, the manifold-related operations  (such as metric evaluation, retraction, and vector transport) in these methods
 account for only a small fraction of the overall computational time.

However, for the solids ($\mathrm{Cu, LiF, Mg, MgO, NaCl}$, and $\mathrm{SiC}$) tested, 
the computational time required grows rapidly  
with respect to the number of $\bk$-points. 
In contrast, the SCF method is less sensitive to the number of $\bk$-points.
In this regard, the SCF method is more efficient than the RCG and RBFGS methods.
We anticipate that the efficiency and robustness of direct minimization methods
can be further enhanced through continous effort on refinement of the implementation. 



\section{Discussions and Conclusion}
\label{sec:discussions}
Direct minimization method on the complex Stiefel manifold in Kohn-Sham density functional theory  
is formulated to treat both finite and extended systems in a unified manner. 
Utilizing the product of Stiefel manifolds, 
we have demonstrated the feasibility of direct minimization calculations with a line search method
for both finite and extended systems. 
In our pilot implementation of the conjugate gradient method 
and tentative version of the BFGS method on 
the complex Stiefel manifold within a compact basis set, 
we conducted comparison tests to reveal advantage and disadvantages of the
Riemannian methods.

Without invoking any preconditioning in the manifold mimimization method, 
we show it can deal with both finite systems compared to the standard SCF method. 
In fact, for finite systems with $\Gamma$-point calculation, it is not necessary to use complex Stiefel manifold. 
Reverting back to the real case, further speed-up is guaranteed by the dimension reduction.
Unfortunately, it is rather slow for extended systems. 
The slow convergence problem for periodic systems should be alleviated with better preconditioning. 
Building upon the linear algebra formulation, 
the implementation of other second-order Riemannian optimization methods, 
such as trust-region methods, may potentially compete against the SCF method.

Currently, the default retraction and vector transport methods are projection-based. 
The main computational bottleneck for the KSDFT problem lies in the evaluation of the objective 
function and its gradient. For a compact basis set, the performance degradation is negligible 
even when using exact geodesic retraction and parallel transport.
However, for future implementations involving a non-compact basis (such as a plane-wave basis), 
geodesic retraction may become increasingly demanding due to the necessity of evaluating 
the matrix exponential. Similarly, the recent Exponential Transformation Direct Minimization (ETDM) method \cite{Ivanov2021} might also face challenges under these conditions.
In such scenarios, the economical retraction and vector transport methods will demonstrate their 
superiority, offering a more efficient alternative without compromising performance.
It is worth pointing out that in the matrix exponential such as ETDM, the number of elements in the exponent  
to be optimized is $M(M+1)$ for 
each $k$-point, much larger than this work, which is $2MN$ when $M\gg N$. Normally, the associated operations scales as 
$O(MN^2)$, much faster than $O(M^3)$.

The framework laid out in this work offers potential conveniences 
for various other electronic structure problems with orthogonality constraints. 
For instance, the self-interaction corrected functional can be directly tested. 
Currently, we are actively investigating the implementation of reduced density matrix functional, 
where additional degrees of freedom, such as the natural occupation number, 
need to be optimized. 
For these computationally intensive objectives, simultaneous optimization should be 
employed to achieve more efficient calculations.
Further research is needed to determine which method—iterative diagonalization or direct minimization—offers greater 
efficiency in RDMFT for periodic systems.




\section*{Acknowledgement}
\label{sec:ack}
K. Luo and T.G. Wang were funded by the National Natural Science Foundation of China under Grant No. 12104230. 
X.G. Ren was funded by the National Natural Science Foundation of China 
under Grant No. 12134012, 12374067, and 12188101.

The authors express their gratitude to Tao Yan and Shimin Zhao for 
their invaluable explanations of nuanced details in Riemannian 
optimization. 
The valuable guidance provided by Daye Zheng on the data structure of ABACUS 
has been immensely beneficial. His insights have greatly enhanced 
our understanding and efficiency in working with the package, 
contributing significantly to the overall progress of our project. 
Additionally, they thank Michael Ulbrich and his 
colleagues for providing the modified KSSOLV code for ensemble DFT 
calculations. The authors are also grateful to Alan S. Edelman, 
Ross A. Lippert, Steven   T. Smith, and Eric J. Bylaska for their generous insights. This work has greatly benefited from the insightful comments and suggestions of anonymous referees.

\clearpage
\appendix
\section{Riemannian BFGS method}
\label{sec:appendix_bfgs}
We present the structure of the RBFGS algorithm\cite{huang2015,huang2018} in Algorithm \ref{algs:bfgs}, where the
search direction is obtained by 
\begin{equation}
    d_k=-B_k^{-1}\grad f(X_k),
\end{equation}
with $B_k^{-1}$ being a linear operator on $T_{X_{k}} \mathrm{St}$, which approximates the 
action of the inverse Hessian along the gradient $\grad f(X_k)$. The symmetry 
and positive definiteness of $B_{k+1}$\cite{huang2018} are ensured by 
\begin{equation}
    B_{k+1}=
    \begin{cases}
    \tilde{B}_{k}-\frac{\tilde{B}_{k}s_{k}(\tilde{B}_{k}^{*}s_{k})^{\dagger}}{\left\langle\tilde{B}_{k}^{*}s_{k},s_{k}\right\rangle}+\frac{y_{k}y_{k}^{\dagger}}{\left\langle y_{k},s_{k}\right\rangle}, & \frac{\left\langle y_{k},s_{k}\right\rangle}{\left\|s_{k}\right\|^{2}}\geq\vartheta(\parallel\grad f(X_{k})\parallel), \\
    \tilde{B}_{k} & \text{otherwise,} 
    \end{cases}
\end{equation}
where  $\tilde{B}_k = \mathcal{T}_{\alpha_k d_k }\circ B_k \circ \left(\mathcal{T}_{\alpha_k d_k }\right)^{-1}$, 
$y_k = \beta_k^{-1} \grad f(X_{k+1}) - \mathcal{T}_{\alpha_k d_k } \left( \grad f(X_k)\right)$, 
$s_k = \mathcal{T}_{\alpha_k d_k }\left( \alpha_k d_k\right)$, $\beta_k = \frac{\| \alpha_k d_k \|}{\| \mathcal{T}^{R}_{\alpha_k d_k } \left(\alpha_k d_k\right) \|}$, 
$\vartheta$ is a function that strictly increasing at 0 and satisfying $\vartheta(0)=0$, 
for instance, we can set $\vartheta(t)=10^{-4}t$. 
Here, the symbol $\circ$ denotes operator composition(or operation composition), meaning that two operations are applied sequentially to an object\cite{ring2012,Absil2008_book}. 
$\mathcal{T}$ is vector transport by projection that its inverse is equal to its adjoint,  
$\mathcal{T}^{R}$ denotes the vector transport by differentiated retraction. Let $A$ be a linear operator on $T_X \mathrm{St}$, $A^*$ denotes the adjoint operator of $A$. 
When the retraction is 
based on the QR decomposition, for any 
$Z, U \in T_X \mathrm{St}$, $\mathcal{T}^{R}$ is\cite{Absil2008_book}
\begin{equation}    
\begin{split}
    \mathcal{T}^{R}_{Z} \left(U\right) &= \mathcal{R}_X(Z) \rho_{\text{skew}} \left( \mathcal{R}_X(Z)^\dagger U \mathcal{R}_X(Z)^\dagger (X + Z)^{-1} \right) 
    \\& \quad + \left( I - \mathcal{R}_X(Z) \mathcal{R}_X(Z)^\dagger \right) U \left( \mathcal{R}_X(Z)^\dagger (X + Z)^{-1} \right),
\end{split}
\end{equation}
\begin{equation}
    (\rho_{\text{skew}}(A))_{ij} = 
    \begin{cases}
    A_{ij}, & \text{if } i > j, \\
    0, & \text{if } i = j, \\
    -A_{ji}, & \text{if } i < j,
    \end{cases}
\end{equation}
where $\mathcal{R}_X (\cdot)$ is the QR decomposition retraction in Eq.\ref{eq:qr_decomp}, $X\in \mathrm{St}$.

In the algorithm,  the inverse Hessian approximation $H_k=B_k^{-1}$ is used instead of $B_k$, and the update formula for $H_k$\cite{huang2018} is
\begin{equation}
    H_{k+1} = \tilde{H}_k - \frac{ \tilde{H}_k y_k, \left(\tilde{H}_k^{*} y_k\right)^\dagger}{\langle \tilde{H}_k^{*} y_k, y_k \rangle} + \frac{s_k s_k^\dagger}{\langle s_k, y_k \rangle}, \quad 
    \tilde{H}_k = \mathcal{T}_{\alpha_k d_k } \circ H_k \circ \left(\mathcal{T}_{\alpha_k d_k }\right)^{-1},
    \label{eq:inverse_Hessian}
\end{equation}
thus, $d_k=-H_k \grad f(X_k)$, and computing $H_k$ is easier than 
calculating the inverse of $B_k$.  $H_i$ and $\tilde{H}_i$ refer to the approximation to the inverse Hessian matrix, 
not the Hamiltonian matrix.  
Meanwhile, one has to choose the initial inverse Hessian approximation $H_0$.
The widely used one is the scaled identity matrix 
\begin{equation}
    H_0 = \gamma I
\end{equation}
where $\gamma$ is a positive scalar and $I$ is the identity matrix. 
One can either utilize gradient information or problem specific estimate of Hessian to 
initialize $H_0$. In this work, we used the most na\"ive choice of $\gamma=1$. 
For multiple $\bk$-points, analogous extension to Algorithm \ref{algs:cg} should be applied.  

\begin{algorithm}
    \caption{BFGS method for minimizing $f(X)$ on the Stiefel manifold}
    \label{algs:bfgs}
    \begin{algorithmic}[1]
    \State Initialization: choose $X_0 \in \mathrm{St}$,  $\epsilon_g, \epsilon_f > 0$,
    $k_{\max}$, 
      $g_0 = \grad f(X_0)$,
      initial inverse Hessian approximation $H_0 = I$ that is symmetric positive definite with respect to the
      metric in Eq.\ref{eq:canonical_metric}
    \While{$\|g_k\| > \epsilon_g$ (or $|f_{k+1} - f_{k}| >  \epsilon_f$) \textbf{and} $k < k_{\max}$}
        \State Compute a  direction as $d_{k} \gets - H_{k}g_{k}$
        \State Line search to find step size $\alpha_k > 0$, 
        and update the point $X_{k+1} \gets  \mathcal{R}_{X_k}( \alpha_k d_k) $ using Eq.~\ref{eq:qr_decomp}  
        \State Compute new Riemannian gradient $g_{k+1} \gets \grad f(X_{k+1})$ 
        and  inverse Hessian approximation $H_{k+1}$ using Eq.~\ref{eq:inverse_Hessian}
        \State $k \gets k + 1$
    \EndWhile
    \end{algorithmic}
\end{algorithm}

Since the retraction based on the QR decomposition and vector transport $\mathcal{T}$ do not satisfy the
locking condition\cite{huang2015}
\begin{equation}
    \mathcal{T}_{\xi}\left(\xi\right)=\beta \mathcal{T}^{R}_{\xi}\left(\xi\right),\ \beta=\frac{\parallel\xi\parallel}{\parallel \mathcal{T}^{R}_{\xi}\left(\xi\right) \parallel},
\end{equation}
the vector transport $\mathcal{T}$ needs to be modified as 
\begin{equation}
    \mathcal{T}_{d}\left(\xi\right)=\left(I-\frac{2\nu_{2}\nu_{2}^{\dagger}}{{\langle \nu_{2}, \nu_{2} \rangle}} \right) \left(I-\frac{2\nu_{1}\nu_{1}^{\dagger}}{{\langle\nu_{1}, \nu_{1}\rangle}}\right)\mathcal{T}_{d}\left(\xi\right),
\end{equation}
where $\nu_{1}=\xi_{1}-\omega,\nu_{2}=\omega-\xi_{2},\xi_{1}=\mathcal{T}_{d}\left(d\right),\xi_{2}=\beta \mathcal{T}^{R}_{d}\left(d\right),Y=\mathcal{R}_{X}(d)$, 
$d$ denotes search direction, $\xi \in T_X \mathrm{St}$, $\omega$ could be any vector in tangent space $T_Y \mathrm{St}$  that satisfies $\| \omega \|=\|\xi_1\|=\|\xi_2\|$, and we take $\omega=-\xi_1$.
Of course, there are other methods to ensure the locking condition\cite{huang2015}. In our computation, we use the modified vector transport to replace the original one. 
Although $\tilde{H}_k = \mathcal{T}_{\alpha_k d_k } \circ H_k \circ \left(\mathcal{T}_{\alpha_k d_k }\right)^{-1}$ is theoretically preferred, our calculations indicate that using $\tilde{H}_k = \mathcal{T}_{\alpha_k d_k } \circ H_k \circ \left(\mathcal{T}_{\alpha_k d_k }\right)^{-1}$ 
to compute the approximate inverse Hessian matrix results in a significant increase 
in the number of iterations and time consumption compared to directly using $\tilde{H}_k=H_k$. 
Additionally, for systems that cannot converge to the SCF calculation results with the 
latter approach, the former approach also does not lead to convergence. Thus, we adopted  $\tilde{H}_k=H_k$.

\section{Flowchart of the SCF method}
\label{sec:appendix_scf}
To fully illustrate the difference between the SCF method and proposed RCG method, a flowchart of the SCF method is attached, which can be compared against Figure \ref{fig:flowchart}.
\begin{figure}[htbp]
    \centering
    \includegraphics[width=0.6\textwidth]{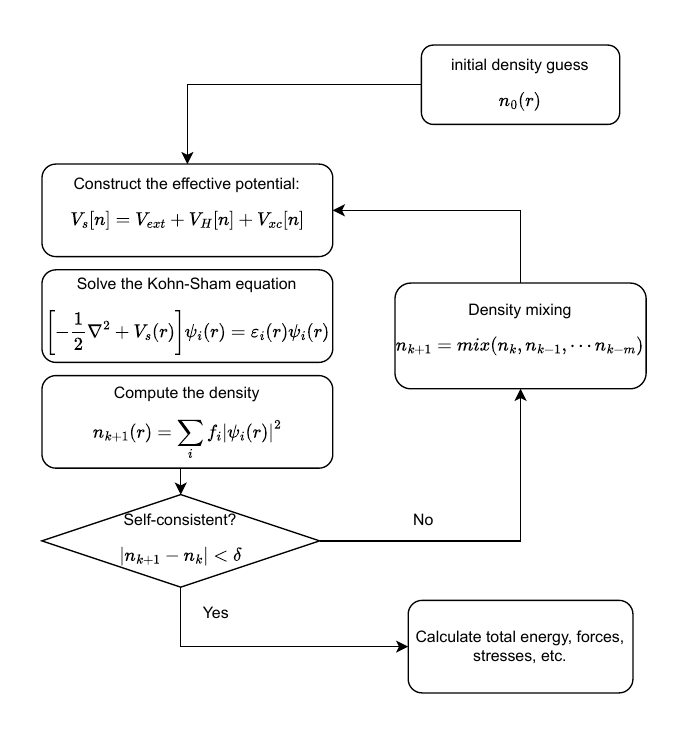}
    \label{fig:ratio}
    \caption{The flowchart for the SCF method.  Different  
    convergence criteria can be chosen.}
    \label{fig:SCF_flowchart}
\end{figure}









\bibliographystyle{elsarticle-num}

\end{document}